\documentclass[a4paper,11pt]{article}
\usepackage{pos}
\usepackage{hyperref}
\usepackage{lineno}

\title{tilepy: rapid tiling strategies in mid/small FoV observatories}
 \ShortTitle{tilepy}

\author*[a]{Fabian Sch\"ussler}
\author[b]{H. Ashkar}
\author[a]{M. de Bony de Lavergne}
\author[c]{M. Seglar-Arroyo}

\affiliation[a]{IRFU, CEA, Université Paris-Saclay, Gif-sur-Yvette, France}
\affiliation[b]{Laboratoire Leprince-Ringuet, Ecole Polytechnique, CNRS, Institut Polytechnique de Paris, Palaiseau, France}
\affiliation[c]{Institut de Física d'Altes Energies (IFAE), Barcelona Institute of Science and Technology, E-08193 Barcelona, Spain}

\emailAdd{fabian.schussler@cea.fr}

\abstract{
The challenges inherent to time-domain multi-messenger astronomy require strategic actions so that adapted, optimized follow-up observations are performed efficiently. In particular, poorly localized events require dedicated tiling and/or targeted, follow-up campaigns so that the region in which the source really is can be efficiently covered, increasing the chances to detect the multi-wavelength counterpart. We have developed the python package "tilepy" to rapidly derive the observation scheduling of large uncertainty localization events by small/mid-FoV instruments. We will describe several mature follow-up scheduling strategies. These range from an option to use of low-resolution grids, to the full integration of sky regions and targeted observations using galaxy catalogs. The algorithms consider the visibility constraints of customisable observatories and allow to schedule observations in both astronomical darkness and in moonlight conditions. Developed initially to provide a rapid response to gravitational wave (GW) alerts by Imaging Atmospheric Cherenkov Telescopes (IACTs), they have been proven successful, as shown by the GW follow-up during O2 and O3 with the H.E.S.S. telescopes, and particularly in the follow-up of GW170817, the first binary neutron star (BNS) merger ever detected. Here we will present a generalisation of these rapid strategies to other alerts showing large uncertainties in the localization, like Gamma-Ray Burst (GRB) alerts from Fermi-GBM. We will also demonstrate the flexibility of {\it tilepy} in scheduling observations for a large variety of observatories. We will conclude by describing the latest developments of these algorithms that are able to derive optimised follow-up schedules across multiple observatories and networks of telescopes.
}

\ConferenceLogo{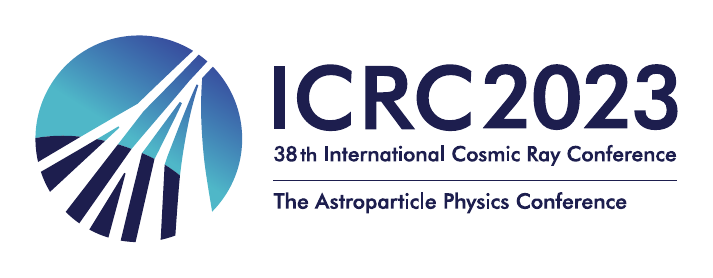}

\FullConference{%
38th International Cosmic Ray Conference (ICRC2023)\\
  26 July - 3 August, 2023\\
  Nagoya, Japan}

\begin{document}
\maketitle

\section{Introduction and main features}
Since 2015, GW detection has been possible with ground-based interferometers like LIGO and Virgo, and this has sparked significant interest in follow-up observations. The main challenge in GW follow-up observations is the poor localization of GW events which can span tens to hundreds of square degrees in the sky, even with advanced GW interferometers.  The localization is provided by probability maps. These maps contain probability information on the localization of the event in the 2D sky, and can also contain information on the distance of the event in each pixel.   To address this, the {\it tilepy} framework presented here comprises various different follow-up strategies.

The following strategies enable to priorize certain regions of the sky according to several reasons, i.e. the astrophysically motivated used of galaxy catalogs and galaxy stellar masses, the optimization of the full coverage of the followup, or the speed of the code. 
The first strategy is a 2D pixel-targeted search, where the highest probability pixel in the localization map is selected for observation. The second strategy is a 2D FoV-targeted search, where the integral probability inside the telescope field of view (FoV) is maximized. {\it tilepy} takes the size of the field-of-view (FoV) of the follow-up observatory into account. Various telescope types provide sizeable FoVs that cover reasonably large regions of the sky in a single exposure. This FoV is given an input parameter to the {\it tilepy} framework and is used as an integration area during the optimization step. The selection of the highest probability position to observe is done with the help of a coordinate grid spatially and temporally correlated with the probability map. The grid displays the positions to be tested for the selection of the highest integrated probability to be observed at a given time. The desired computation speed and calculation accuracy combined with telescope properties such as the FoV play a role in selecting the resolution of the grid. i.e. the number of test positions and the spacing between them. These parameters can be selected by the user. 

The remaining three strategies are based on 3D approaches that consider the distribution of galaxies in the local universe and the estimation of the distance of the event in each pixel of the map. These strategies are developed for events for which distance information is provided by the gravitational wave detectors. The probability from the map is convolved with the galaxy distribution to reduce the search area. Each galaxy in the area is assigned a probability of hosting the event. The user can also choose to take galaxy stellar masses into account in the assignment of the probability following~\cite{MANGROVE}. The third approach is a 3D galaxy-targeted search. Galaxies are targeted individually. The galaxy with the highest probability of hosting the event is selected for observation. The fourth strategy is a 3D FoV-targeted search where the probability of the galaxies is integrated inside the telescope FoV. This strategy typically favors groups and clusters of galaxies instead of individual galaxies. It is adapted for large-to-medium-Fov-telescopes to optimize coverage. To select the positions to be tested, galaxies play the role of seed positions. The probability is integrated around each galaxy and the highest integrated galaxy probability is selected for observations at a given time. In the case of a large localization region containing a large number of galaxies, the user can choose the fifth strategy that uses a coordinate grid to reduce the computation time. Like in the 2D case, the user-defined resolution of the grid can be adjusted for quick computation. A detailed explanation of these five strategies can be found in~\citep{2021JCAP...03..045A}. 

In all cases, observation and visibility conditions are taken into consideration. The positions selected for observation for a given time are masked for the next iteration of probability computation. The telescope observation and visibility parameters, the allocated observation duration, and the telescope parameters (position, FoV, maximum zenith angle) are defined by the user. Moreover, the user can define threshold cuts. For example, a follow-up observation is triggered only if a certain probability, defined by the user is covered. Finally, further optimization of the observations can favor observations at low airmass (i.e. low zenith angles) to reduce absorption effects that could influence the data quality and instrument performance (e.g. the energy threshold of VHE gamma-ray observatories). This is done by assigning a higher weight (chosen by the user) for decreasing zenith angle observations.

\begin{figure}[h!]
\begin{center}
\includegraphics[width= 0.85\textwidth]{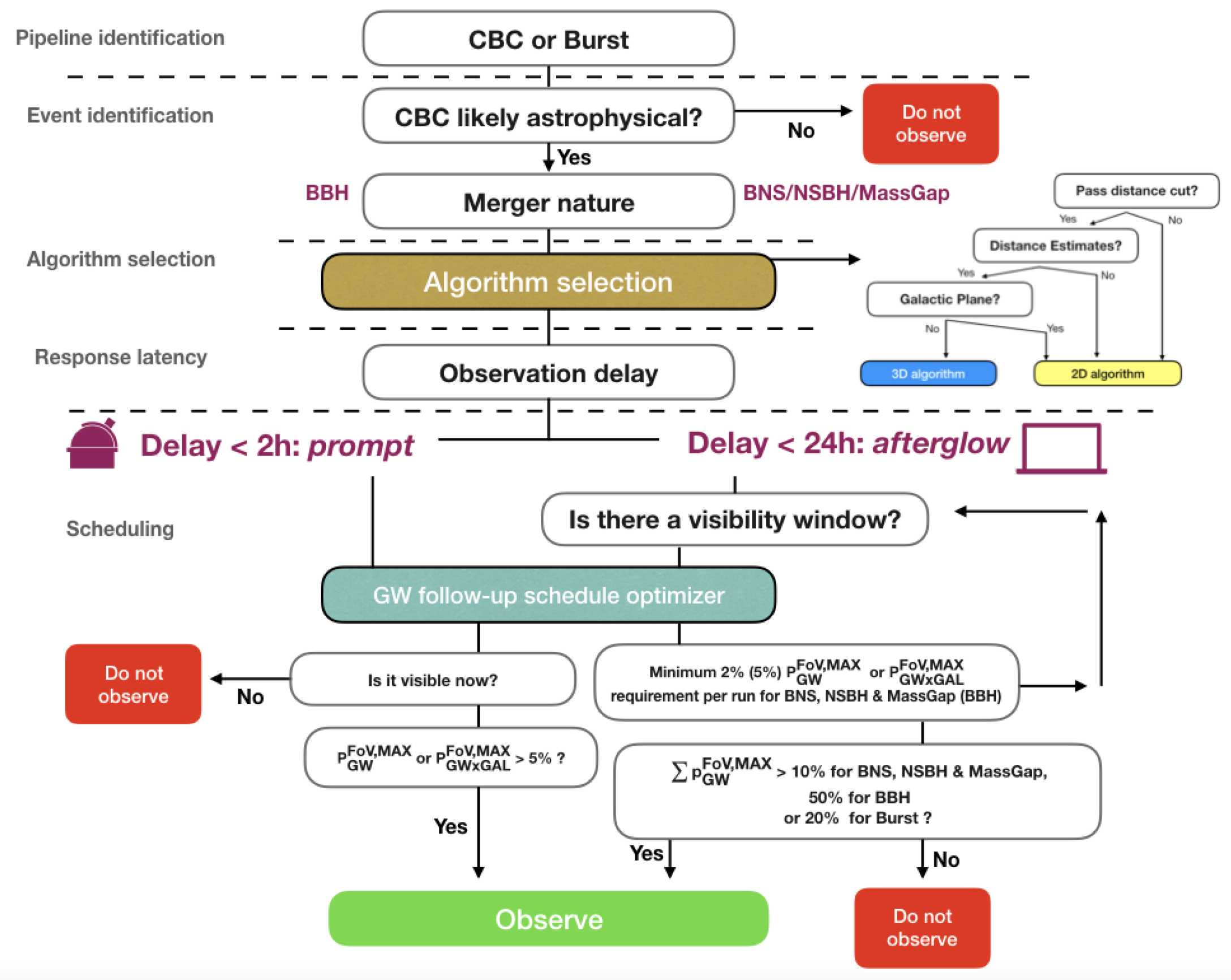}
\caption{Schematic overview of the {\it tilepy} decision tree used in the automatic response of H.E.S.S to GW events during O3. From~\citep{2021JCAP...03..045A}}
\label{fig:hess}
\end{center}
\end{figure}

To automate the follow-up process, the various options are integrated into the {\it tilepy} framework. Upon reception of a GW alert, it selects the appropriate strategy (2D or 3D, and the mentioned sub-categories) based on the available event-by-event information. A 2D strategy is favored for distant events and events falling behind the galactic plane where galaxy catalog completion is low. The thresholds and settings for these selections can be defined by each user and for each use case. The criteria used by the H.E.S.S. collaboration during the third LIGO/Virgo observing run O3 are given in Fig.~\ref{fig:hess}.

\section{Use case examples}
We here briefly describe a few use cases of {\it tilepy} by scientific collaborations and multi-messenger platforms. They have been made possible largely thanks to the publication of the {\it tilepy} source code. It is available (as beta version) on \url{https://github.com/astro-transients/tilepy}.

\subsection{GW follow-up observations with very-high-energy gamma-ray instruments}
{\it tilepy} has been developed within and for the H.E.S.S. very-high-energy gamma-ray observatory~\citep{2021JCAP...03..045A, hessTOOsystem}. During the second and third LIGO/Virgo observing runs, H.E.S.S. triggered observations on eight GW events. Observations were obtained for the well-known GW170817. In this case, H.E.S.S. was the first ground-based pointing instrument to observe the location of the merger origin in NGC 4993~\citep{GW170817_HESS}. H.E.S.S. furthermore observed several other events with varying degrees of localization~\cite{2021ApJ...923..109A}. The H.E.S.S. response latency to prompt alerts was found to be less than 1 minute. Further details of these observations are given in~\citep{2021ApJ...923..109A}. An update is presented at this conference~\citep{2023ICRC_HESS_GWs}.

Overall, the H.E.S.S. GW Rapid Follow-up Program demonstrated successful follow-up observations of GW events using a combination of 2D and 3D strategies provided by {\it tilepy}. The program's automated response capabilities and optimized observation scheduling allowed for efficient utilization of H.E.S.S. telescopes during GW observing runs.

The first prototype of the Larged-Telescope Array (LST-1), which is part of the grid of telescopes of the northern site of the Cherenkov Telescope Array Observatory (CTAO), handles alerts via the Transient Handler~\citep{2021arXiv210804309C}. The automatic response to gravitational waves starting in O4 is ensured by {\it tilepy}. The automatic response is currently being tested as the handling of events is different to that presented in Fig \ref{fig:hess}. The {\it tilepy} package has been adapted to study prospects of detections at very-high energy gamma-rays of sGRB as GW counterpart, in future observing runs by CTAO with optimised strategies. More information on this study is presented at this conference \cite{GWatCTA}.  

\subsection{Fermi/GBM follow-up observations with very-high-energy gamma-ray instruments}

The {\it Fermi}/GBM instrument with it's extremely large FoV, provide a lot of GRB alert to the community, but follow-up and search of counterpart to these alerts could be tricky due to the fairly large uncertainty on the position (at least a few degrees, up to a few tens of degrees). 

Since {\bf find date and reference}, {\it Fermi}/GBM alerts provide an uncertainty localisation map with a format close to the ones provided by GW interferometers. {\it tilepy} is able to read this map and generate a schedule for observation. This has been successfully used since 2020 by both H.E.S.S. and LST-1 experiments to optimize {\it Fermi}/GBM follow-ups.

\subsection{tilepy.com}
While the {\it tilepy} framework is publicly available and can be run by anyone on a local machine, a centraly maintained and easy to use interface enables much more diverse use cases. The {\it tilepy} development team partnered with the Astro-COLIBRI multi-messenger platform to provide a cloud-based computing environment with open access to the scheduling calculation. This new service is running at \url{https://tilepy.com}. 

The central component is a Python-based Flask API, which provides the {\it tiling} endpoint. It wraps a large number of the {\it tilepy} input parameters and provides direct access to the latest tagged version of {\it tilepy}. The various parameters are documented on \url{https://tilepy.com/apidoc}, which also provides a ready to use example.

\subsection{Astro-COLIBRI}
Astro-COLIBRI is a general platform that aims to simplify the analysis and follow-up observations of transient astrophysical phenomena by providing interactive and user-friendly graphical tools, while seamlessly integrating real-time detections from multiple messengers.

The {\it tilepy} API described above has also been integrated into the Astro-COLIBRI platform. Using the Astro-COLIBRI user interfaces on the web and as smartphone apps, the user can easily request the calculation of an observation plan from the {\it tilepy} API. The result is displayed in the form of a table and fully integrated into the Astro-COLIBRI platform. Each observation position and the related telescope FoV is illustrated on the sky maps to allow for rapid identification of additional multi-messenger information within the covered regions. The total coverage achieved with the proposed observation plan is mentioned to allow for a rapid decision making process towards follow-up observations.

\begin{figure}
\centering
    \begin{minipage}{0.75\textwidth}
        \centering
        \includegraphics[width=0.9\textwidth]{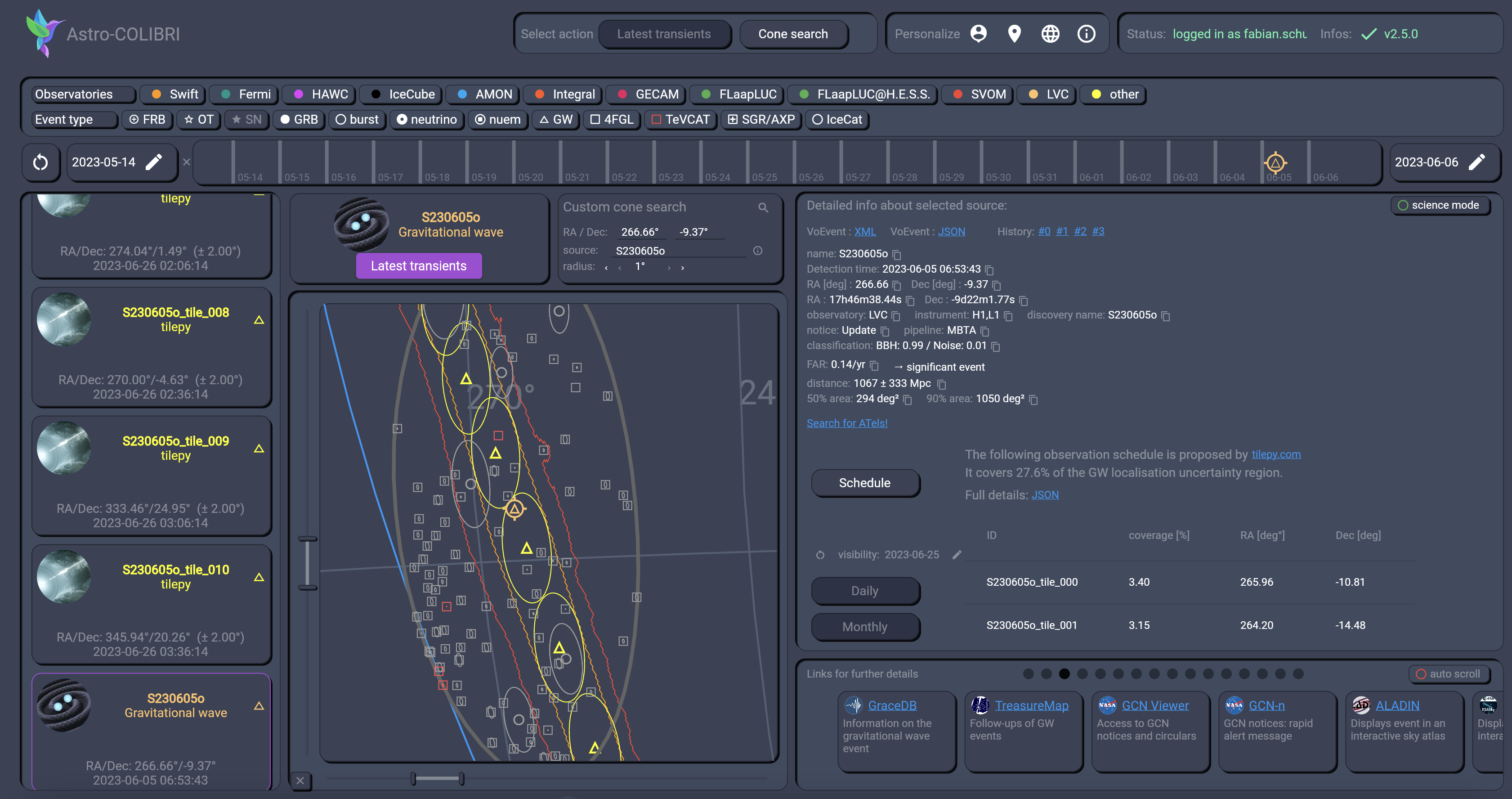} 
        \caption{Tilepy observation plan displayed in the Astro-COLIBRI web frontend (\url{https:astro-colibri.com})~\citep{2023ICRC_Astro-COLIBRI} }\label{fig:astro-colibri_web}
    \end{minipage}\hfill
    \begin{minipage}{0.2\textwidth}
        \centering
        \includegraphics[width=\textwidth]{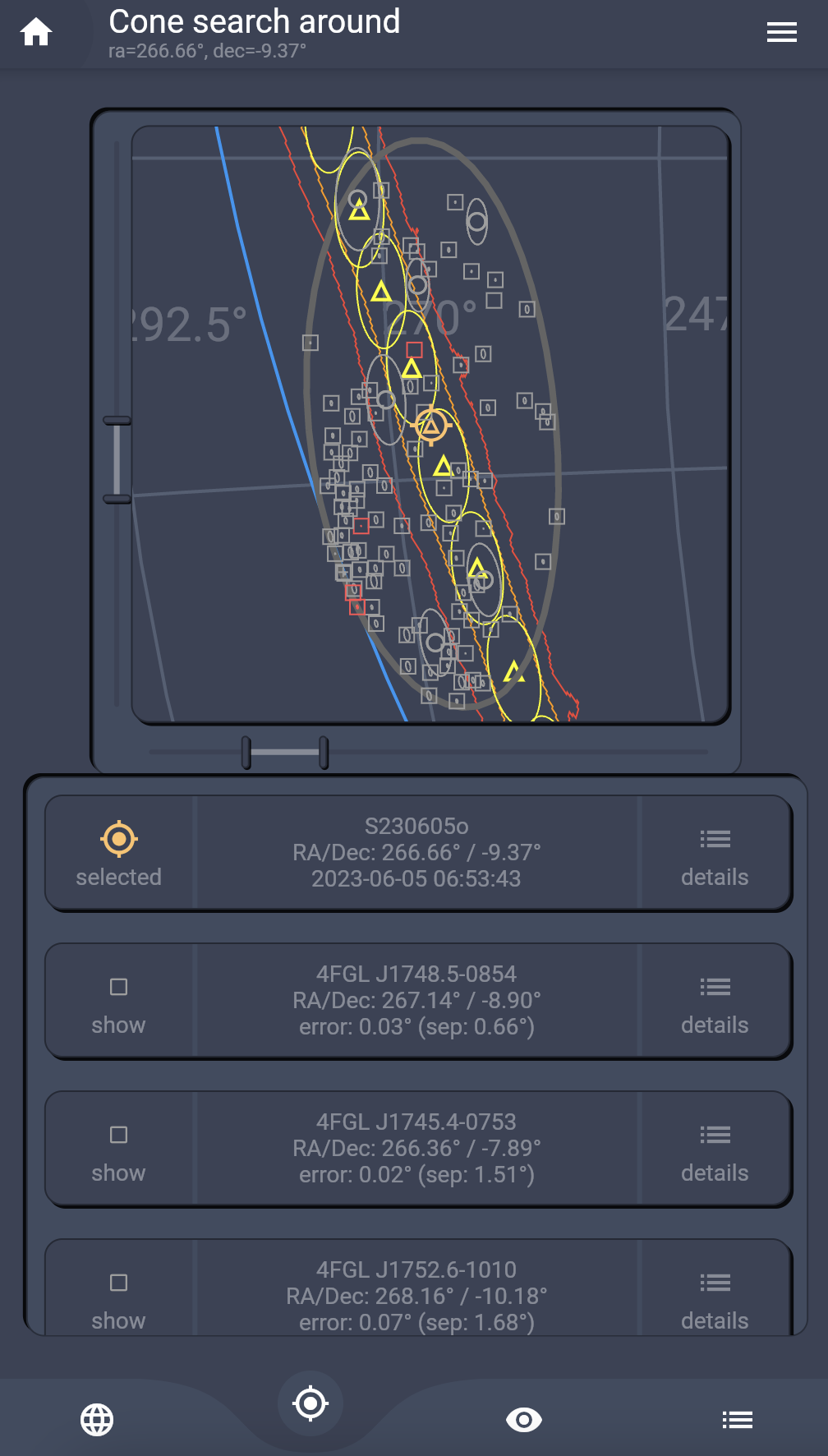} 
        \caption{Smartphone view}\label{fig:astro-colibri_app}
    \end{minipage}
\end{figure}

\section{Conclusions and outlook}
The {\it tilepy} package is a powerful tool for scheduling follow-up observations of GW events. It provides a variety of strategies for optimizing the coverage of the localization region, taking into account the properties of the follow-up telescope and the visibility conditions. The code is open source, which makes it accessible to a wide range of researchers.

The package is still under development, and there are a number of planned improvements. These include the addition of new strategies for optimizing the observation scheduling, and the continued development of the Astro-COLIBRI interface for easy use. Here are some specific areas where the {\it tilepy} package could be improved:
\begin{itemize}
    \item Optimisation and re-factoring of the {\it tilepy} code base and use of more up-to-date libraries for certain tasks like the handling of the multiorder healpix maps providing the localisation information.
    \item The addition of new strategies for optimizing the observation scheduling. For example, the code could be extended to include strategies that take into account the expected brightness of the counterpart, the sensitivity of the telescope, and the time since the GW event.
    \item Further integration into the Astro-COLIBRI multi-messenger platform. This would allow for the {\it tilepy} package to be used to schedule follow-up observations of GW events in conjunction with other data, such as gamma-ray bursts or neutrinos.
\end{itemize}

We believe that the {\it tilepy} package has the potential to be a valuable tool for multi-messenger follow-up observations. With continued, community driven, development, it could become an essential tool for maximizing the chances of detecting a counterpart to a GW event.

\section{Acknowledgements}
The authors acknowledge the support of the French Agence Nationale de la Recherche (ANR) under reference ANR-22-CE31-0012. This work was also supported by the Programme National des Hautes Energies of CNRS/INSU with INP and IN2P3, co-funded by CEA and CNES.  MSA acknowledges the support of Grant FJC2020-044895-I funded by MCIN/AEI/10.13039/501100011033 and by the European Union NextGenerationEU/PRTR. 

\bibliography{tilepy_ICRC2023}{}
\bibliographystyle{aasjournal} %


%

%

\end{document}